\documentclass{main}
\usepackage{amsmath}
\usepackage{amssymb}
\usepackage{textcomp}
\usepackage{fancyhdr}
\usepackage{lastpage}
\usepackage{subfig}

\usepackage[numbers,sort&compress]{natbib}
\setcitestyle{square,nonamebreak}

\def\jpsi{J/\psi}

\def\be{\begin{equation}}
\def\ee{\end{equation}}
\def\bea{\begin{eqnarray}}
\def\eea{\end{eqnarray}}

\newcommand{\Photo}{\includegraphics[height=35mm]{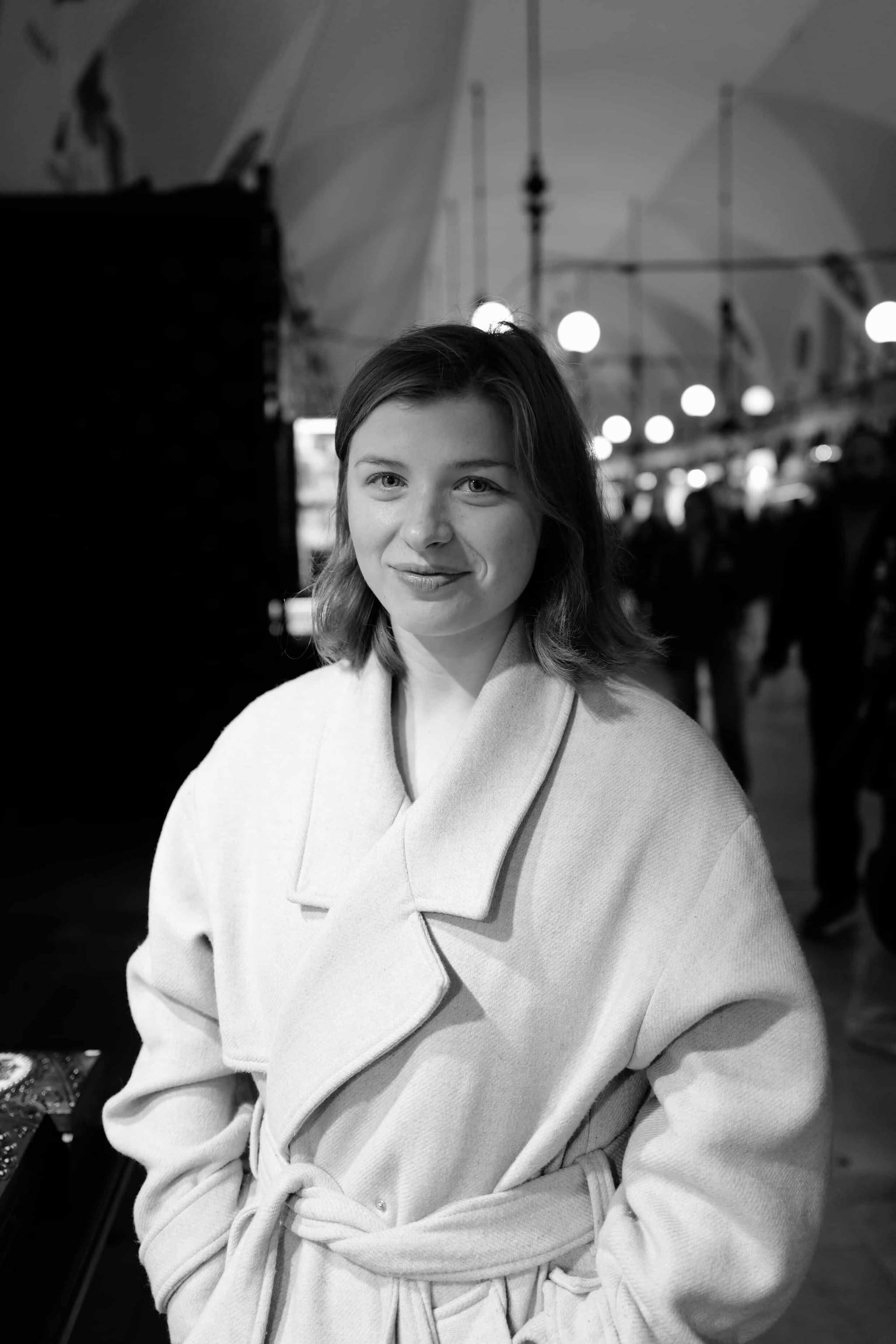}}

\fancypagestyle{firstpagefooter}{%
  \fancyfoot[L]{  \textcopyright \footnotesize This work is an open access article under a Creative Commons Attribution 4.0 International License (https://creativecommons.org/licenses/by/4.0/).}
  \fancyfoot[C]{}  
   
}

\begin{document}

\thispagestyle{firstpagefooter}
\title{\Large Inclusive quarkonium photoproduction selection and the effect of pileup at the LHC}

\author{\underline{K. Lynch}$^{1,2}$ \footnote{kate.lynch1@ucdconnect.ie}, J.P. Lansberg$^{1}$, R.~McNulty$^{2}$, C.~Van Hulse$^{3}$}

\address{
$^{1}$Universit\'e Paris-Saclay, CNRS, IJCLab, 91405 Orsay, France\\
$^{2}$School of Physics, University College Dublin, Dublin 4, Ireland\\
$^{3}$University of Alcal\'{a}, 28801 Alcal\'{a} de Henares (Madrid), Spain}

\maketitle\abstracts{
Measurements of inclusive quarkonium photoproduction provide strong constraints on the quarkonium production mechanism; however, this process has not yet been measured at the LHC. We summarise our previously developed selection strategy for isolating inclusive quarkonium photoproduction in $p$Pb collisions at the LHC, which offer an optimal balance of photon flux, luminosity, and low pileup. We further examine the applicability of our selection criteria in different collision systems. While our method can be readily extended to PbPb collisions, $pp$ collisions require additional care due to the significantly higher pileup. We discuss how pileup affects the selection criteria, highlighting that it substantially degrades forward-detector–based selections and reduces the efficiency of rapidity-gap requirements.
}



\section{Introduction}\label{sec:introduction}
Quarkonia are bound states of heavy quarks, either charm--anti-charm or bottom--anti-bottom. They are in principle the simplest hadrons. Yet, there is no theoretical description of their production that can describe all of the {experimental} data. Owing to certain computational simplifications and distinct kinematical dependencies {and despite being statistically limited}, inclusive photoproduction data from HERA (see refs~\cite{ZEUS:2012qog,H1:2010udv} and therein) provide strong constraining power when combined with the highly precise LHC hadroproduction data~\cite{Lansberg:2019adr}. {These also offer a potential for gluon PDFs measurement~\cite{Jung:1992uj,ColpaniSerri:2021bla}}.
This calls for more inclusive quarkonium photoproduction data. {This will naturally come with the future US EIC~\cite{Boer:2024ylx}. However, as we have argued recently~\cite{Lansberg:2024zap}, such data can also be collected at the LHC.}

The isolation of photoproduction in hadron-hadron collisions is performed through the selection of ultra-peripheral collisions (UPCs). These are defined as interactions mediated over distances larger than the sum of the radii of the colliding hadrons. At such impact parameters, the {colliding hadrons do not overlap, QCD exchanges are suppressed}, and photon-induced interactions dominate.

So far, quarkonium photoproduction measurements from the LHC have focused on exclusive/diffractive processes~\cite{ALICE:2021tyx,ALICE:2018oyo,ALICE:2012yye,CMS:2018bbk,LHCb:2021bfl,LHCb:2021hoq,ALICE:2014eof,LHCb:2022ahs,ALICE:2013wjo,ALICE:2019tqa,ALICE:2021gpt,CMS:2016itn,LHCb:2013nqs,LHCb:2014acg,LHCb:2018rcm,LHCb:2015wlx,ATLAS:2017sfe,ATLAS:2015wnx,CMS:2011vma}, which are measurable via fully determined final states.\footnote{Photon-induced production of $J/\psi$ has also been isolated in peripheral, as opposed to ultra-peripheral, PbPb collisions~\cite{ALICE:2022zso,ALICE:2015mzu,LHCb:2021hoq}, where an excess of low-$P_T$ $\jpsi$ with respect to the expected hadroproduction yield was observed.} However, inclusive measurements offer much more insight into the poorly understood production mechanism, {in particular when hadronisation is accompanied by radiation}. In~\cite{Lansberg:2024zap}, we demonstrated the feasibility of making a measurement of inclusive quarkonium photoproduction at the LHC. To extract the inclusive photoproduction cross section, we detailed a set of event-selection criteria tailored to the specific capabilities of the LHC experiments. At the time we made this study, there were only one {final and one on-going} inclusive photoproduction study from LHC data~\cite{ATLAS:2021jhn,ATLAS:2022cbd}. Since then there have been many new final and preliminary studies~\cite{ATLAS:2024mvt,ATLAS:2025tof,ATLAS:2025hac,CMS:2025jjx}, which highlights the novelty of this area of research.

Our analysis~\cite{Lansberg:2024zap} focused on $p$Pb collisions at the LHC as the Pb ion is the most probable photon emitter. {On the contrary,} in $pp$ and PbPb collisions, one must address the ambiguity in the identity of the photon source. In addition, in $pp$ collisions, photoproduction cross sections are much smaller than in $p$Pb collisions due to the reduced size of the proton photon flux. This reduction can be compensated by the much higher luminosity recorded in LHC $pp$ collisions. However, as we will discuss {in this proceedings contribution,} the much higher pileup present in $pp$ collisions constrains the selection criteria that can be used.


\section{Inclusive photoproduction selection requirements}\label{sec:selection}

\begin{figure}[hbt!]
    \centering
    \subfloat[]{\includegraphics[width=0.25\linewidth]{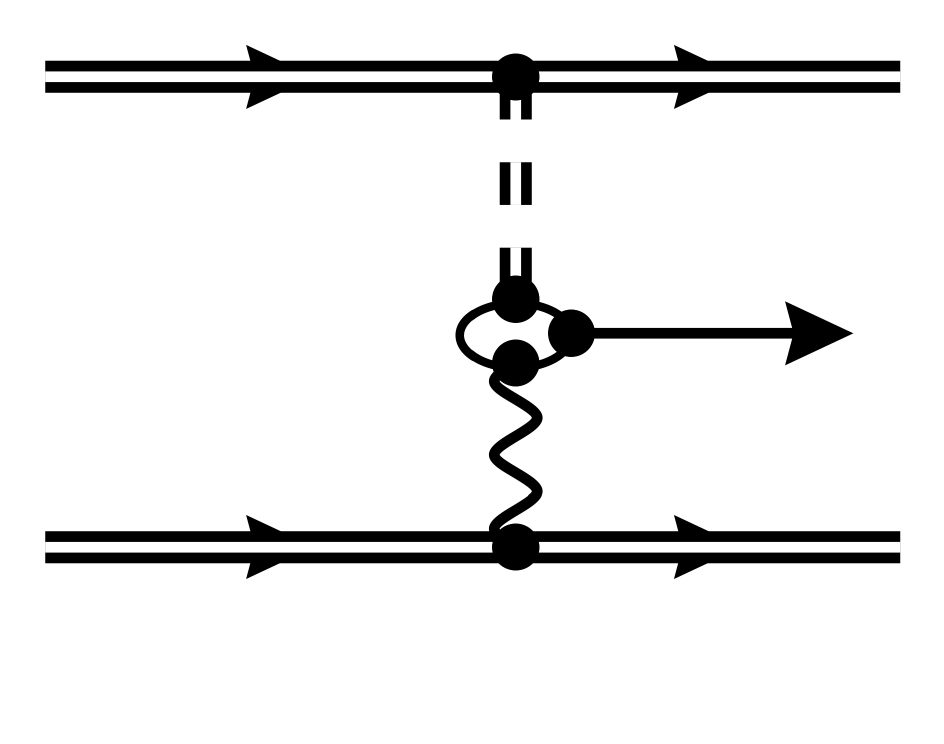}}
    \subfloat[]{\includegraphics[width=0.25\linewidth]{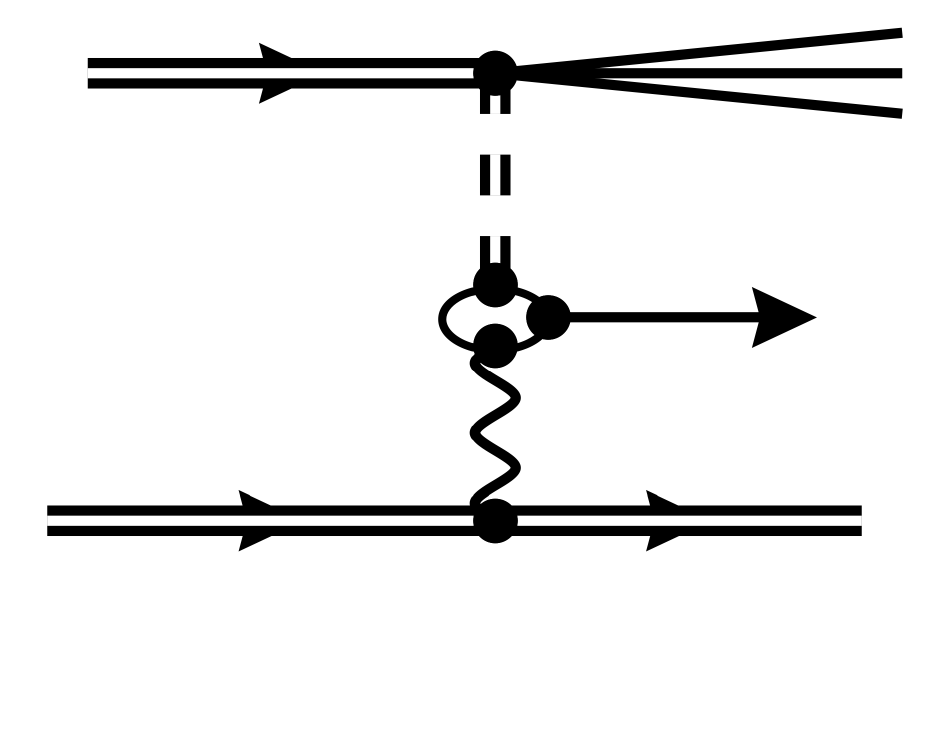}}
    \subfloat[]{\includegraphics[width=0.25\linewidth]{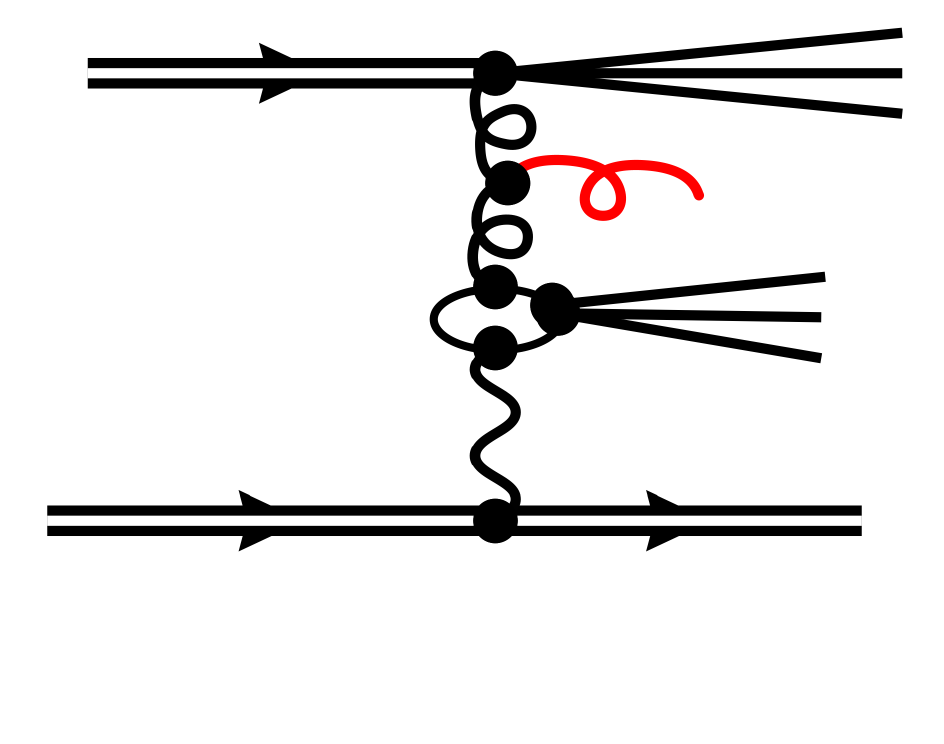}}
    \subfloat[]{\includegraphics[width=0.25\linewidth]{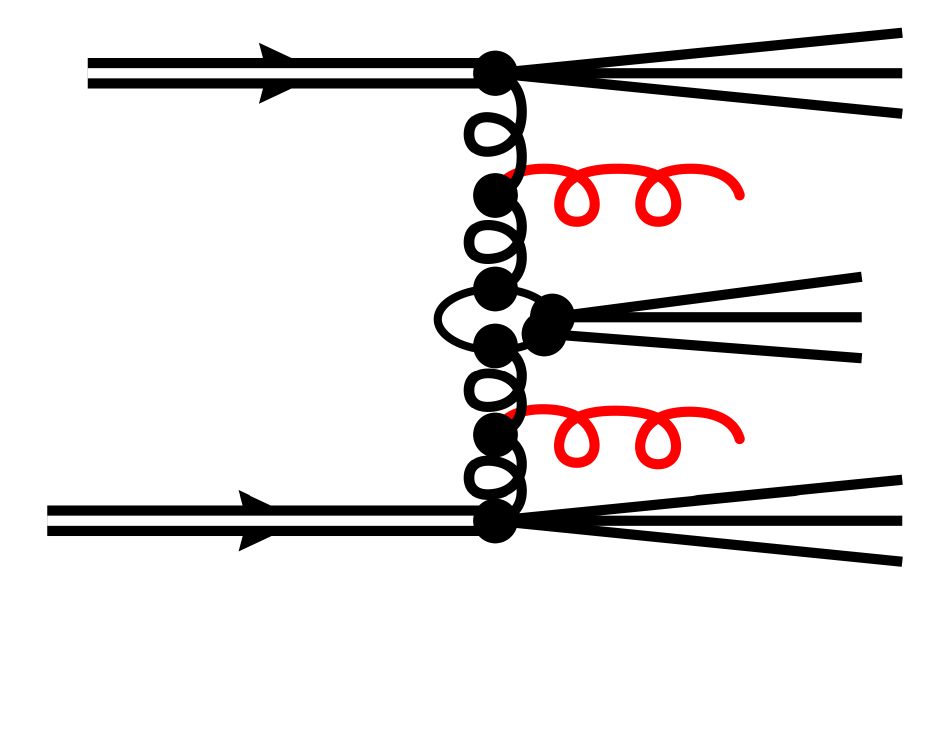}}
    \caption{(a) exclusive diffractive photoproduction; (b)  diffractive photoproduction with nuclear break up; (c) inclusive photoproduction; (d) inclusive hadroproduction. Figures are produced using FeynGame~\cite{Bundgen:2025utt}.}
    \label{fig:graphs}
\end{figure}

As mentioned, photon-induced quarkonium measurements from the LHC to date consist of diffractive processes, i.e., photon–Pomeron fusion. These processes involve a colourless exchange between the colliding hadrons, resulting in the exclusive production of a quarkonium state and no other particles. Figure~\ref{fig:graphs} (a) illustrates the fully exclusive diffractive process, while Fig.~\ref{fig:graphs}(b) shows the diffractive process accompanied by nuclear break-up. In both cases, event selection relies on a combination of characteristic criteria:
\begin{itemize}
    \item Central exclusivity — detection of a quarkonium state with a veto on additional activity;
     \item Colourless exchange — detected as an absence of particle activity between the beam particles and the produced system;
     \item Intact photon emitter — which can be identified using various detector signatures.

\end{itemize}

For inclusive quarkonium photoproduction (see Fig.~\ref{fig:graphs}(c)), these selection criteria must be relaxed to allow for an inclusive central system and a coloured exchange on one side of the interaction. Relaxing the diffractive criteria introduces a major experimental challenge: the presence of large hadronic backgrounds that dominate the signal (see Fig.~\ref{fig:graphs}(d)). For $J/\psi$ and $\Upsilon$ photoproduction, the hadronic background is typically $\mathcal{O}(100-10,000)$ times larger than the photoproduction cross section, depending on the kinematic region~\cite{Lansberg:2024zap}. Consequently, this is an experimentally demanding measurement. This difficulty is not unique to quarkonium, but applies broadly to inclusive photoproduction measurements, which remain a relatively new experimental domain.

 In what follows we outline the UPC selection criteria discussed in ref.~\cite{Lansberg:2024zap}. 

\subsection{Centrality and zero degree calorimeters}

The centrality of a hadron-hadron collision characterises the degree of hadronic overlap between the colliding nuclei.\footnote{In what follows, we globally refer to nucleons or nuclei as hadrons.} The most central collisions correspond to nearly head-on interactions, whereas the most peripheral--or ultra-peripheral--collisions involve little to no nuclear overlap.

The photoproduction contribution within a given event sample can be enhanced by restricting the analysis to the most peripheral centrality classes. Centrality can be determined in various ways; one common method is by measuring the energy deposited in a far-forward detector, such as a Zero Degree Calorimeter (ZDC). The ZDCs are installed on both sides of the interaction point in ALICE, ATLAS, and CMS and are sensitive to neutral particles with $\eta \gtrsim 8$. The main source of these neutral particles is neutrons coming from Pb ions that are broken during hadronic collisions. However, neutrons passing through the ZDCs can also come from the photonuclear excitation and de-exciation of an ion, resulting in the emission of one or more neutrons. 

In an event sample, the relative fraction of photoproduction processes increases as one selects less central collisions. Based on Ref.~\cite{ALICE:2015kgk}, we estimate that the 80–100\% centrality class contains only about 6\% of {the total amount of} hadronically produced 
$\jpsi$ events, {which is the dominant experimental background}. Therefore, selecting this centrality class—previously measured for $\jpsi$ production in proton–lead collisions~\cite{ALICE:2015kgk}—is expected to suppress approximately 94\% of the hadroproduction background.

{In ref.~\cite{Lansberg:2024zap}, we demonstrated that by selecting the 80–100\% centrality class, in combination with rapidity gap requirements that are discussed in the following subsection, it is possible to measure the photoproduction cross section up to $P_T = 20$~GeV at the LHC. We note that the resolution of the ZDC detectors is such that they can resolve single to few neutron emissions. Recent analyses~\cite{ATLAS:2024mvt,ATLAS:2025tof,CMS:2025jjx} have used this fact to define a $0nXn$ selection criterion, where there are no neutrons present on the photon-going side and at least one on the side with hadronic break-up. In ref.~\cite{Lansberg:2024zap} we did not have a reliable way to model forward neutron emissions and so conservatively used the 80–100\% centrality class. However, using a tighter selection criterion than the 80–100\% centrality class, such as $0nXn$, is expected to significantly enhance the signal purity.}

Following methods described in \cite{Baltz:2002pp,Broz:2019kpl}, we compute the cross sections and median impact parameters, $\text{med}(b)$, for the inclusive photoproduction of $J/\psi$ in $p$Pb collisions at 8.16~TeV and PbPb collisions at 5.12~TeV. These results appear in Table~\ref{tb:sigmaresults}. It can be seen that in $p$Pb collisions the probability for a photoproduced $J/\psi$ accompanied by at least one neutron is $\mathcal{O}(0.01\%)$. Thus, requiring no neutron emissions has an extremely high efficiency for retaining signal. However, in PbPb collisions we estimate this same probability to be $\mathcal{O}(20\%)$. This cut for PbPb is less efficient and it biases the photon flux to larger impact parameters and should be handled with care.

\begin{table}[h!]
\caption{Inclusive $\jpsi$ photoproduction cross section, 
its fractional contribution to the total cross section, and the median impact parameter, $\text{med}(b)$, in $p$Pb and PbPb collision systems with different requirements on forward neutron emissions. }
\begin{center}
\begin{tabular}{ |c|ccc| } 
\hline
 & $\sigma$ & \% of total &$\text{med}(b)$  \\
\hline
\multicolumn{4}{|c|}{$p$ Pb $\rightarrow$ Pb+$X$n $\oplus$ $J/\psi$ $X$  at $\sqrt{s_{NN}}=8.16$~TeV}\\

\hline
total & 55~$\mu$b & 100\% & 41~fm \\
$0n$ & 55~$\mu$b & 99.99\% & 41~fm \\
$1n$ & 3~nb & 0.005\% & 11~fm \\
$Xn$ &  7~nb & 0.01\% & 11~fm \\

\hline

\multicolumn{4}{|c|}{PbPb $\rightarrow$ Pb+$X$n $\oplus$ $J/\psi$ $X$ at $\sqrt{s_{NN}}=5.12$~TeV}\\

\hline
total & 12~mb & 100\% & 64~fm \\
$0n$ & 10~mb & 82\% & 92~fm \\
$1n$ & 0.6~mb & 5\% & 23~fm \\
$Xn$ &  2~mb & 18\% & 21~fm \\
 \hline
\end{tabular}
\end{center}
\label{tb:sigmaresults}
\end{table}

\subsection{Rapidity gaps}\label{sec:RG}
For inclusive photoproduction, one expects a region devoid of particle activity—known as a rapidity gap—between the centrally produced system and the photon-emitting hadron. This gap arises from the colourless nature of the photon exchange involved in the process.

There is, however, some flexibility in how a rapidity gap is defined. Depending on the analysis, one may adopt either a stricter or looser definition, for example by varying the minimum required size of the gap or the threshold for vetoing particle activity. Moreover, the performance of the rapidity-gap selection depends strongly on the detector pseudorapidity coverage: a broader coverage allows for a more effective identification of rapidity gaps and, consequently, a more efficient background suppression.

{The strictest form of a rapidity gap requires no activity within a specified pseudorapidity region, whereas a looser gap definition allows for some limited particle production or radiation in that region. In ref.~\cite{Lansberg:2024zap}, we explored the strictest type of gap within the detector coverages of ALICE, ATLAS, CMS, and LHCb. Among these, ATLAS and CMS provide the broadest continuous pseudorapidity coverage, of order $\mathcal{O}(10\text{~units})$, which grants them superior discrimination power for identifying rapidity gaps. For ALICE, our study was limited to the central barrel region ($|\eta| < 0.8$), and thus we underestimated the selection potential of ALICE. The inclusion of the V0 detectors in the analysis, as employed in ref.~\cite{ALICE:2023mfc}, would extend the pseudorapidity coverage and improve the ALICE rapidity-gap capability.}

We defined $\Delta \eta_\gamma \equiv \min\{|\eta_{\gamma\text{-edge}}-\eta_i |\}$ for $i\neq \jpsi$, where $\eta_{\gamma\text{-edge}}$ is the pseudorapidity of the edge of the detector on the photon-going side and $\eta_i$ are the pseudorapidities of particles within the detector acceptance, which for CMS corresponds to a charged track acceptance of $|\eta_\text{cal}|<2.5$, $p_{T\text{,cal}}>0.2$~GeV and a calorimeter acceptance of $2.5<|\eta_\text{cal}|<4.9$, $p_{T\text{,cal}}>0.2$~GeV. 

\begin{figure}[hbt!]
        \centering
\subfloat[]{\label{fig:standardgap}\includegraphics[width=0.5\textwidth]{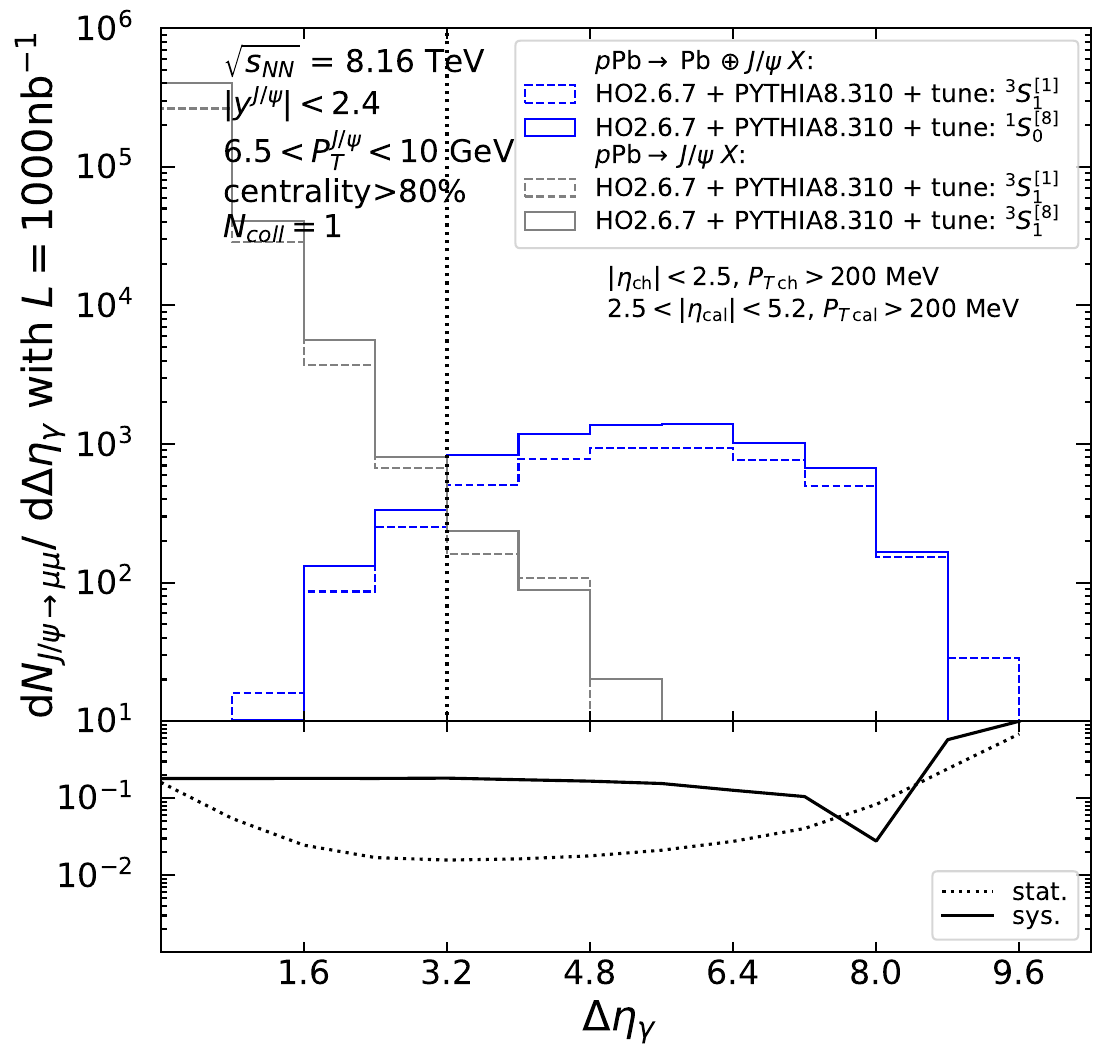}}\subfloat[]{\label{fig:sumofgaps}\includegraphics[width=0.5\textwidth]{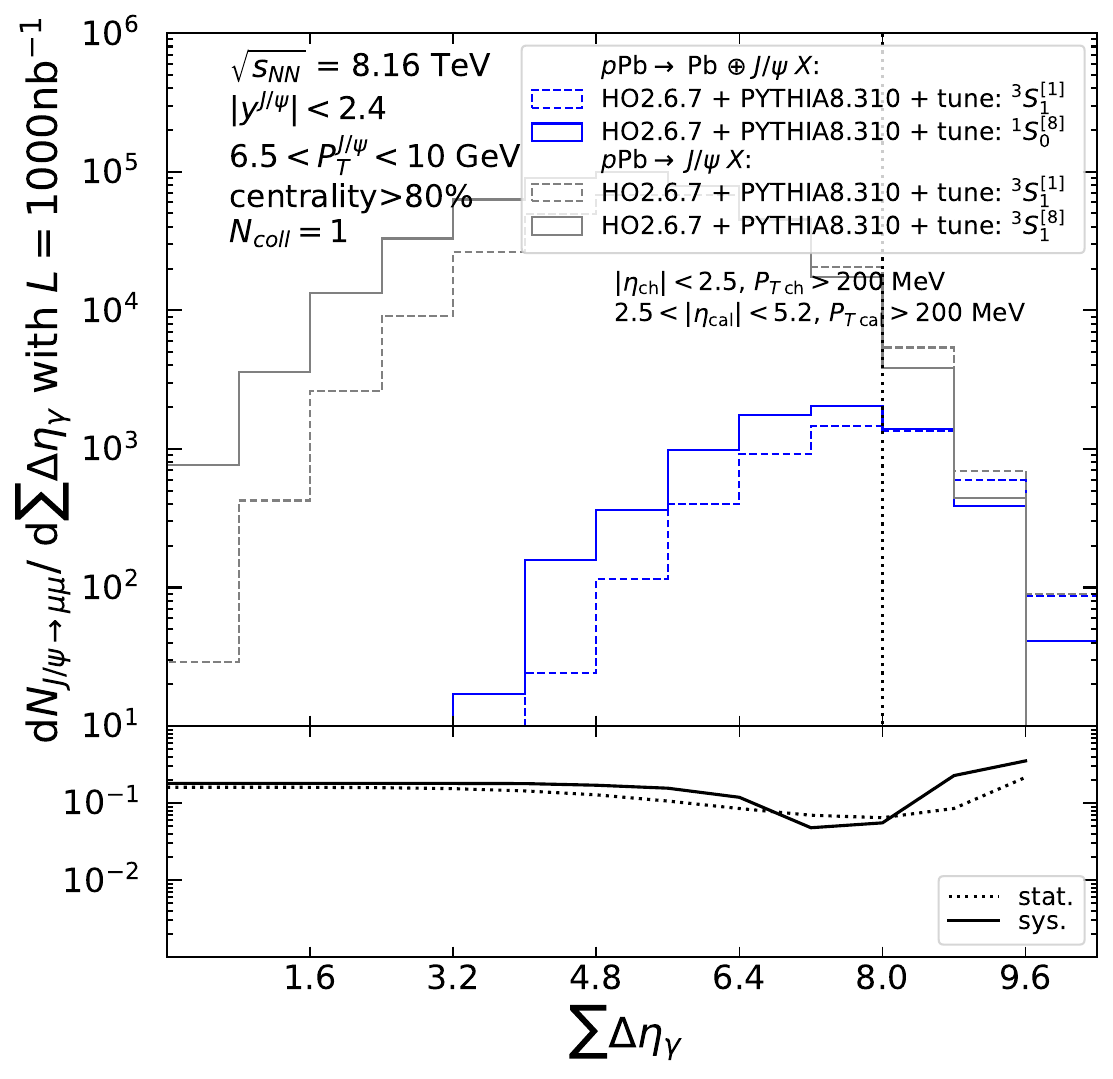}}
      
    \caption{\small Comparison between (a) $\Delta\eta_\gamma$ and (b) $\sum\Delta\eta_\gamma$ for $J/\psi$ using the singlet (dashed) and octet (solid) tunes of direct photoproduction (blue) and hadroproduction (grey) in the peripheral limit with $N_\text{coll}=1$. The lower panel shows the relative statistical (dotted) and systematic (solid) uncertainties as a function of the cut value on $\Delta \eta_\gamma$ or $\sum\Delta\eta_\gamma$. 
    The dotted vertical line indicates the cut value that minimises the statistical uncertainty following the background subtraction.  }
 \label{fig:CMS-rapgap-Ncoll1}
\end{figure}

A looser cut was introduced in ATLAS UPC analyses~\cite{ATLAS:2021jhn,ATLAS:2022cbd}: a \emph{cumulative rapidity-gap} criterion, which evaluates the pseudorapidity separation between pairs of adjacent particles and sums all intervals where $\Delta\eta > 0.5$. A comparison between the single-gap ($\Delta\eta_\gamma$) and cumulative-gap ($\sum\Delta\eta_\gamma$) distributions is shown in Fig.~\ref{fig:CMS-rapgap-Ncoll1} for direct photoproduction (blue) and hadroproduction (grey) in the peripheral limit\footnote{The peripheral limit corresponds to the 90-100\% centrality class; these events are $pp$-like with $N_\text{coll}=1$, where $N_\text{coll}$ denotes the number of binary nucleon–nucleon collision.}. The cumulative-gap definition exhibits {a reduced efficiency both for rejecting hadroproduction background and for retaining the photoproduction signal} compared to the single-gap requirement. {As shown in Fig.~2, the statistical uncertainty in the selected sample is reduced by approximately a factor of four when using the cumulative-gap definition relative to the single-gap one.} In our modelling of the photoproduction signal (blue) and the hadroproduction background (grey), we used both octet and singlet channels for each, generated with \texttt{HELAC-Onia+PYTHIA}~\cite{Shao:2012iz,Shao:2015vga,Bierlich:2022pfr,Sjostrand:2014zea}, and tuned to data; see Ref.~\cite{Lansberg:2024zap} for details. 

\subsection{HeRSCheL at LHCb}\label{sec:Hersh}

The High-Rapidity Shower Counters for LHCb (HeRSCheL) detector was installed in the LHCb experiment during Run~2. Although it was removed for Run~3, there are ongoing internal discussions within LHCb regarding its possible reinstatement for Run~4. The system consisted of five plastic scintillator panels sensitive to charged-particle showers in the very forward and backward regions, $5 < |\eta| < 10$~\cite{Akiba_2018}. Activity observed in these regions is indicative of beam-particle break-up; therefore, a veto on HeRSCheL activity on one side of the muon arm should efficiently select photoproduced events while rejecting hadroproduction backgrounds. In ref.~\cite{Lansberg:2024zap}, we demonstrated the qualitative difference between hadroproduction and photoproduction {particle activity within the HeRSCheL pseudorapidity acceptance. We showed that photoproduction events exhibit no particle activity in this region, whereas hadroproduction events show considerable activity.} However, due to the lack of detailed modelling in the far-forward region, we did not make a quantitative statement regarding this effect.
\section{Pileup and its effect on the above methods}

Pileup is defined as the number of simultaneous collisions per bunch crossing and is characterized by a Poisson distribution with mean $\langle \mu \rangle$. An important advantage of $p$Pb and PbPb collisions compared to $pp$ collisions is the significantly reduced pileup. To illustrate this, we quote the Run~2 conditions reported by CMS~\cite{CMS:2023yxy}:

\begin{itemize}
    \item at $\sqrt{s_{NN}} = 8.16$~TeV, $\langle \mu_{p\text{Pb}} \rangle = 0.25$, $\mathcal{L}_\text{int}=1.7\times 10^2$~nb$^{-1}$,
    \item at $\sqrt{s_{NN}} = 5.02$~TeV, $\langle \mu_{\text{PbPb}} \rangle = 0.0004$, $\mathcal{L}_\text{int}=1.7$~nb$^{-1}$,
    \item at $\sqrt{s_{NN}} = 5.02$~TeV, $\langle \mu_{pp} \rangle = 3$, $\mathcal{L}_\text{int}=3.0\times 10^5$~nb$^{-1}$,
\end{itemize}
where the $pp$ run at 5.02~TeV was a Pb-ion reference run, and $\mathcal{L}_\text{int}$ is the integrated luminosity. In typical high-luminosity Run 2 $pp$ collisions, $\langle \mu \rangle$ is of order 30~\cite{CMS_DP20020}.  

Assuming the following scaling for the photoproduction cross sections 
\[
\sigma^\gamma_{pp} : \sigma^\gamma_{p\text{Pb}} : \sigma^\gamma_{\text{PbPb}} \propto 1 : Z^2 : Z^2 A, 
\] 
and the hadroproduction cross sections 
\[
\sigma^h_{pp} : \sigma^h_{p\text{Pb}} : \sigma^h_{\text{PbPb}} \propto 1 : A : A^2,
\] 
where $A=208$ and $Z=82$. Based on the above luminosities and scalings, we anticipate similar numbers of photoproduction events, $N=\sigma \times \mathcal{L}_\text{int}$, in each collision system and an enhanced signal to background ratio in $p$Pb and PbPb collisions compared to $pp$ collisions. 

For $\langle \mu \rangle = 0.25$, the expected ratio of single-interaction events ($n = 1$) to multi-interaction events ($n > 1$) is approximately 8:1, allowing for high event retention. When $\langle \mu \rangle \sim 1$, some degradation in analysis efficiency occurs, with the ratio of single- to multi-interaction events falling to roughly 1.4:1. For higher pile-up values ($\langle \mu \rangle \gtrsim 1$), the applicability and effectiveness of the methods discussed above must be carefully reconsidered.

The performance of the HeRSCheL detector is affected by pileup, as it measures the total charged energy in the forward region during each beam collision~\cite{Akiba:2018neu}. Consequently, it cannot distinguish multiple interactions occurring within a single bunch crossing, leading to a reduction in its discriminating power under high pile-up conditions. Similarly, the ZDC lacks the timing precision to disentangle events in the presence of pileup~\cite{Suranyi:2021ssd}. 

On the other hand, rapidity gaps may still be used. In the presence of multiple collisions in a single beam crossing, energy deposits in calorimeters originating from different collision vertices cannot always be disentangled. Therefore, rapidity-gap selections must be restricted to the tracking region, where collision vertices can be cleanly separated.\footnote{Tight vertex cuts were used by CMS to measure $\gamma \gamma \to \tau^+ \tau^-$ in $pp$ collisions with high pile-up~\cite{CMS:2024qjo}.} Consequently, only selections based on charged-particle tracks are usable. This limitation reduces the accessible pseudorapidity range for experiments such as ATLAS and CMS, from roughly 10 units down to about 5 units in $\Delta \eta$, covering $|\eta| < 2.5$. However, the coverages of LHCb and ALICE used in our simulations~\cite{Lansberg:2024zap} remain unchanged. 

%


%
\section{Conclusion}

We have summarised selection criteria that enable hadron beams to be used as photon beams for the study of inclusive processes. The techniques discussed here may be generalised to many processes. We have discussed how the run conditions, in particular the pileup, play a role in the efficiency of the selection of inclusive photoproduction events. The same may be said of photoproduction events in general. We emphasise that an inclusive quarkonium photoproduction measurement at the LHC is highly anticipated and will shed light on the production mechanism.


\section*{Acknowledgments}
{\it Funding --- }The research conducted in this publication was funded by the Irish Research Council under grant number GOIPG/2022/478.
This project has also received funding from the Agence Nationale de la Recherche (ANR) via the grants ANR-20-CE31-0015 (``PrecisOnium''), 
ANR-24-CE31-7061-01 (``3DLeaP"), and via the IDEX Paris-Saclay ``Investissements d'Avenir" (ANR-11-IDEX-0003-01) through the GLUODYNAMICS project funded by the ``P2IO LabEx (ANR-10-LABX-0038)" and through the Joint PhD Programme of Universit\'e Paris-Saclay (ADI).
This work  was also partly supported by the French CNRS via the IN2P3 projects ``GLUE@NLO" and ``QCDFactorisation@NLO".

\bibliography{UPC-2025-KL}

\begin{thebibliography}{10}

\bibitem{ZEUS:2012qog}
H.~Abramowicz et~al.
\newblock {Measurement of inelastic $J/\psi$ and $\psi^\prime$ photoproduction
  at HERA}.
\newblock {\em JHEP}, 02:071, 2013.

\bibitem{H1:2010udv}
F.~D. Aaron et~al.
\newblock {Inelastic Production of J/psi Mesons in Photoproduction and Deep
  Inelastic Scattering at HERA}.
\newblock {\em Eur. Phys. J. C}, 68:401--420, 2010.

\bibitem{Lansberg:2019adr}
Jean-Philippe Lansberg.
\newblock {New Observables in Inclusive Production of Quarkonia}.
\newblock {\em Phys. Rept.}, 889:1--106, 2020.

\bibitem{Jung:1992uj}
H.~Jung, G.~A. Schuler, and J.~Terron.
\newblock {J / psi production mechanisms and determination of the gluon density
  at HERA}.
\newblock {\em Int. J. Mod. Phys. A}, 7:7955--7988, 1992.

\bibitem{ColpaniSerri:2021bla}
Alice Colpani~Serri, Yu~Feng, Carlo Flore, Jean-Philippe Lansberg, Melih~A.
  Ozcelik, Hua-Sheng Shao, and Yelyzaveta Yedelkina.
\newblock {Revisiting NLO QCD corrections to total inclusive $J/\psi$ and
  $\Upsilon$ photoproduction cross sections in lepton-proton collisions}.
\newblock {\em Phys. Lett. B}, 835:137556, 2022.

\bibitem{Boer:2024ylx}
Dani{\"e}l Boer et~al.
\newblock {Physics case for quarkonium studies at the Electron Ion Collider}.
\newblock {\em Prog. Part. Nucl. Phys.}, 142:104162, 2025.

\bibitem{Lansberg:2024zap}
Jean-Philippe Lansberg, Kate Lynch, Charlotte Van~Hulse, and Ronan McNulty.
\newblock {Inclusive photoproduction of vector quarkonium in ultra-peripheral
  collisions at the LHC}.
\newblock {\em Eur. Phys. J. C}, 85:161, 2025.

\bibitem{ALICE:2021tyx}
Shreyasi Acharya et~al.
\newblock {First measurement of the |$t$|-dependence of coherent $J/\psi$
  photonuclear production}.
\newblock {\em Phys. Lett. B}, 817:136280, 2021.

\bibitem{ALICE:2018oyo}
Shreyasi Acharya et~al.
\newblock {Energy dependence of exclusive $\mathrm {J}/\psi $ photoproduction
  off protons in ultra-peripheral p\textendash{}Pb collisions at
  $\sqrt{s_{\mathrm {\scriptscriptstyle NN}}} = 5.02$ TeV}.
\newblock {\em Eur. Phys. J. C}, 79(5):402, 2019.

\bibitem{ALICE:2012yye}
Betty Abelev et~al.
\newblock {Coherent $J/\psi$ photoproduction in ultra-peripheral Pb-Pb
  collisions at $\sqrt{s_{NN}} = 2.76$ TeV}.
\newblock {\em Phys. Lett. B}, 718:1273--1283, 2013.

\bibitem{CMS:2018bbk}
Albert~M. Sirunyan et~al.
\newblock {Measurement of exclusive $\Upsilon$ photoproduction from protons in
  pPb collisions at $\sqrt{s_\mathrm{NN}} =$ 5.02 TeV}.
\newblock {\em Eur. Phys. J. C}, 79(3):277, 2019.
\newblock [Erratum: Eur.Phys.J.C 82, 343 (2022)].

\bibitem{LHCb:2021bfl}
Roel Aaij et~al.
\newblock {Study of coherent $J/\psi$ production in lead-lead collisions at
  $\sqrt{s_{NN}} = 5 TeV$}.
\newblock 7 2021.

\bibitem{LHCb:2021hoq}
Roel Aaij et~al.
\newblock {$J/\psi$ photoproduction in Pb-Pb peripheral collisions at $\sqrt
  {s_{NN}}$= 5 TeV}.
\newblock {\em Phys. Rev. C}, 105(3):L032201, 2022.

\bibitem{ALICE:2014eof}
Betty~Bezverkhny Abelev et~al.
\newblock {Exclusive $\mathrm{J/}\psi$ photoproduction off protons in
  ultra-peripheral p-Pb collisions at $\sqrt{s_{\rm NN}}=5.02$ TeV}.
\newblock {\em Phys. Rev. Lett.}, 113(23):232504, 2014.

\bibitem{LHCb:2022ahs}
R.~Aaij et~al.
\newblock {Study of exclusive photoproduction of charmonium in ultra-peripheral
  lead-lead collisions}.
\newblock {\em JHEP}, 06:146, 2023.

\bibitem{ALICE:2013wjo}
E.~Abbas et~al.
\newblock {Charmonium and $e^+e^-$ pair photoproduction at mid-rapidity in
  ultra-peripheral Pb-Pb collisions at $\sqrt{s_{\rm NN}}$=2.76 TeV}.
\newblock {\em Eur. Phys. J. C}, 73(11):2617, 2013.

\bibitem{ALICE:2019tqa}
Shreyasi Acharya et~al.
\newblock {Coherent J/$\psi$ photoproduction at forward rapidity in
  ultra-peripheral Pb-Pb collisions at $\sqrt{s_{\rm{NN}}}=5.02$ TeV}.
\newblock {\em Phys. Lett. B}, 798:134926, 2019.

\bibitem{ALICE:2021gpt}
Shreyasi Acharya et~al.
\newblock {Coherent $\rm{J/\psi}$ and $\rm{\psi'}$ photoproduction at
  midrapidity in ultra-peripheral Pb-Pb collisions at
  $\sqrt{s_{\mathrm{NN}}}~=~5.02$ TeV}.
\newblock {\em Eur. Phys. J. C}, 81(8):712, 2021.

\bibitem{CMS:2016itn}
Vardan Khachatryan et~al.
\newblock {Coherent $J/\psi$ photoproduction in ultra-peripheral PbPb
  collisions at $\sqrt {s_{NN}} =$ 2.76 TeV with the CMS experiment}.
\newblock {\em Phys. Lett. B}, 772:489--511, 2017.

\bibitem{LHCb:2013nqs}
R~Aaij et~al.
\newblock {Exclusive $J/\psi$ and $\psi$(2S) production in pp collisions at $
  \sqrt{s} = 7$ TeV}.
\newblock {\em J. Phys. G}, 40:045001, 2013.

\bibitem{LHCb:2014acg}
Roel Aaij et~al.
\newblock {Updated measurements of exclusive $J/\psi$ and $\psi$(2S) production
  cross-sections in pp collisions at $\sqrt{s}=7$ TeV}.
\newblock {\em J. Phys. G}, 41:055002, 2014.

\bibitem{LHCb:2018rcm}
Roel Aaij et~al.
\newblock {Central exclusive production of $J/\psi$ and $\psi(2S)$ mesons in
  $pp$ collisions at $\sqrt{s}=13~$TeV}.
\newblock {\em JHEP}, 10:167, 2018.

\bibitem{LHCb:2015wlx}
Roel Aaij et~al.
\newblock {Measurement of the exclusive \ensuremath{\Upsilon} production
  cross-section in pp collisions at $ \sqrt{s}=7 $ TeV and 8 TeV}.
\newblock {\em JHEP}, 09:084, 2015.

\bibitem{ATLAS:2017sfe}
Morad Aaboud et~al.
\newblock {Measurement of the exclusive $\gamma \gamma \rightarrow \mu^+ \mu^-$
  process in proton-proton collisions at $\sqrt{s}=13$ TeV with the ATLAS
  detector}.
\newblock {\em Phys. Lett. B}, 777:303--323, 2018.

\bibitem{ATLAS:2015wnx}
Georges Aad et~al.
\newblock {Measurement of exclusive $\gamma\gamma\rightarrow \ell^+\ell^-$
  production in proton-proton collisions at $\sqrt{s} = 7$ TeV with the ATLAS
  detector}.
\newblock {\em Phys. Lett. B}, 749:242--261, 2015.

\bibitem{CMS:2011vma}
Serguei Chatrchyan et~al.
\newblock {Exclusive photon-photon production of muon pairs in proton-proton
  collisions at $\sqrt{s}=7$ TeV}.
\newblock {\em JHEP}, 01:052, 2012.

\bibitem{ALICE:2022zso}
Shreyasi Acharya et~al.
\newblock {Photoproduction of low-$p_{\rm T}$ J/$\psi$ from peripheral to
  central Pb$-$Pb collisions at 5.02 TeV}.
\newblock {\em Phys. Lett. B}, 846:137467, 2023.

\bibitem{ALICE:2015mzu}
Jaroslav Adam et~al.
\newblock {Measurement of an excess in the yield of $J/\psi$ at very low
  $p_{\rm T}$ in Pb-Pb collisions at $\sqrt{s_{\rm NN}}$ = 2.76 TeV}.
\newblock {\em Phys. Rev. Lett.}, 116(22):222301, 2016.

\bibitem{ATLAS:2021jhn}
Georges Aad et~al.
\newblock {Two-particle azimuthal correlations in photonuclear ultraperipheral
  Pb+Pb collisions at 5.02 TeV with ATLAS}.
\newblock {\em Phys. Rev. C}, 104(1):014903, 2021.

\bibitem{ATLAS:2022cbd}
{Photo-nuclear jet production in ultra-peripheral Pb+Pb collisions at
  $\sqrt{s}_\text{NN} = 5.02$ TeV with the ATLAS detector}.
\newblock 2022.

\bibitem{ATLAS:2024mvt}
Georges Aad et~al.
\newblock {Measurement of photonuclear jet production in ultraperipheral Pb+Pb
  collisions at sNN=5.02{\,}{\,}TeV with the ATLAS detector}.
\newblock {\em Phys. Rev. D}, 111(5):052006, 2025.

\bibitem{ATLAS:2025tof}
Georges Aad et~al.
\newblock {Charged-hadron and identified-hadron
  (KS0,~{\ensuremath{\Lambda}},~{\ensuremath{\Xi}}{\ensuremath{-}}) yield
  measurements in photonuclear Pb+Pb and p+Pb collisions at sNN=5.02TeV with
  ATLAS}.
\newblock {\em Phys. Rev. C}, 111(6):064908, 2025.

\bibitem{ATLAS:2025hac}
Georges Aad et~al.
\newblock {Characterization of nuclear breakup as a function of hard-scattering
  kinematics using dijets measured by ATLAS in $p$+Pb collisions}.
\newblock 4 2025.

\bibitem{CMS:2025jjx}
Vladimir Chekhovsky et~al.
\newblock {Measurement of D$^0$ meson photoproduction in ultraperipheral heavy
  ion collisions}.
\newblock 9 2025.

\bibitem{Bundgen:2025utt}
Lars B{\"u}ndgen, Robert~V. Harlander, Sven~Yannick Klein, and Magnus~C.
  Schaaf.
\newblock {FeynGame 3.0}.
\newblock {\em Comput. Phys. Commun.}, 314:109662, 2025.

\bibitem{ALICE:2015kgk}
Jaroslav Adam et~al.
\newblock {Centrality dependence of inclusive J/\ensuremath{\psi} production in
  p-Pb collisions at $ \sqrt{s_{\mathrm{NN}}}=5.02 $ TeV}.
\newblock {\em JHEP}, 11:127, 2015.

\bibitem{Baltz:2002pp}
Anthony~J. Baltz, Spencer~R. Klein, and Joakim Nystrand.
\newblock {Coherent vector meson photoproduction with nuclear breakup in
  relativistic heavy ion collisions}.
\newblock {\em Phys. Rev. Lett.}, 89:012301, 2002.

\bibitem{Broz:2019kpl}
M.~Broz, J.~G. Contreras, and J.~D. Tapia~Takaki.
\newblock {A generator of forward neutrons for ultra-peripheral collisions:
  n00n}.
\newblock {\em Comput. Phys. Commun.}, 253:107181, 2020.

\bibitem{ALICE:2023mfc}
{Exclusive and dissociative J/$\psi$ photoproduction, and exclusive dimuon
  production, in p$-$Pb collisions at $\sqrt{s_{\rm NN}} = 8.16$ TeV}.
\newblock 4 2023.

\bibitem{Shao:2012iz}
Hua-Sheng Shao.
\newblock {HELAC-Onia: An automatic matrix element generator for heavy
  quarkonium physics}.
\newblock {\em Comput. Phys. Commun.}, 184:2562--2570, 2013.

\bibitem{Shao:2015vga}
Hua-Sheng Shao.
\newblock {HELAC-Onia 2.0: an upgraded matrix-element and event generator for
  heavy quarkonium physics}.
\newblock {\em Comput. Phys. Commun.}, 198:238--259, 2016.

\bibitem{Bierlich:2022pfr}
Christian Bierlich et~al.
\newblock {A comprehensive guide to the physics and usage of PYTHIA 8.3}.
\newblock 3 2022.

\bibitem{Sjostrand:2014zea}
Torbj\"orn Sj\"ostrand, Stefan Ask, Jesper~R. Christiansen, Richard Corke,
  Nishita Desai, Philip Ilten, Stephen Mrenna, Stefan Prestel, Christine~O.
  Rasmussen, and Peter~Z. Skands.
\newblock {An introduction to PYTHIA 8.2}.
\newblock {\em Comput. Phys. Commun.}, 191:159--177, 2015.

\bibitem{Akiba_2018}
K.~Carvalho Akiba, F.~Alessio, N.~Bondar, W.~Byczynski, V.~Coco, P.~Collins,
  R.~Dumps, R.~Dzhelyadin, P.~Gandini, B.R.~Gruberg Cazon, R.~Jacobsson,
  D.~Johnson, J.~Manthey, J.~Mauricio, R.~McNulty, S.~Monteil, B.~Rachwal,
  M.~Ravonel Salzgeber, L.~Roy, H.~Schindler, S.~Stevenson, and G.~Wilkinson.
\newblock The {HERSCHEL} detector: high-rapidity shower counters for {LHCb}.
\newblock {\em Journal of Instrumentation}, 13(04):P04017--P04017, apr 2018.

\bibitem{CMS:2023yxy}
{Performance of CMS muon reconstruction in heavy ion collisions}.
\newblock 2023.

\bibitem{CMS_DP20020}
{CMS Collaboration}.
\newblock Cms public results dp20020.
\newblock TWiki page, CERN, 2020.

\bibitem{Akiba:2018neu}
K.~Carvalho Akiba et~al.
\newblock {The HeRSCheL detector: high-rapidity shower counters for LHCb}.
\newblock {\em JINST}, 13(04):P04017, 2018.

\bibitem{Suranyi:2021ssd}
O.~Sur\'anyi et~al.
\newblock {Performance of the CMS Zero Degree Calorimeters in pPb collisions at
  the LHC}.
\newblock {\em JINST}, 16(05):P05008, 2021.

\bibitem{CMS:2024qjo}
Aram Hayrapetyan et~al.
\newblock {Observation of $\gamma\gamma\to\tau\tau$ in proton-proton collisions
  and limits on the anomalous electromagnetic moments of the $\tau$ lepton}.
\newblock {\em Rept. Prog. Phys.}, 87(10):107801, 2024.

\end{thebibliography}

\end{document}